%% file: main.tex
\documentclass[runningheads]{llncs}
\usepackage[T1]{fontenc}
\usepackage{graphicx}
\usepackage{xcolor}
\usepackage{multirow}
\usepackage{amsmath}
\usepackage{cite}
\begin{document}
\institute{}
\title{From low resource information extraction to identifying influential nodes in knowledge graphs\thanks{\small DISTRIBUTION STATEMENT A. Approved for public release. Distribution is unlimited. \hfill\break \textcopyright 2024 MASSACHUSETTS INSTITUTE OF TECHNOLOGY.\hfill \break
Delivered to the U.S. Government with Unlimited Rights, as defined in DFARS Part 252.227-7013 or 7014 (Feb 2014). Notwithstanding any copyright notice, U.S. Government rights in this work are defined by DFARS 252.227-7013 or DFARS 252.227-7014 as detailed above. Use of this work other than as specifically authorized by the U.S. Government may violate any copyrights that exist in this work.\hfill\break
This material is based upon work supported by the Department of Defense under Air Force Contract No. FA8702-15-D-0001. Any opinions, findings, conclusions or recommendations expressed in this material are those of the authors and do not necessarily reflect the views of the Department of Defense.}}

\titlerunning{Low resource information extraction}
%
\author{Erica Cai \and Olga Simek\and
Benjamin A. Miller \and Danielle Sullivan-Pao \and Evan Young \and
Christopher L. Smith}
%
\authorrunning{E. Cai et al. }
%
\institute{MIT Lincoln Laboratory, Lexington MA 02421, USA\\
\and
University of Massachusetts Amherst, Amherst MA 01003, USA\\
\email{ecai@cs.umass.edu}}
%
\maketitle              
\begin{abstract}
We propose a pipeline for identifying important entities from intelligence reports that constructs a knowledge graph, where nodes correspond to entities of fine-grained types (e.g.~\textit{traffickers}) extracted from the text and edges correspond to extracted relations between entities (e.g.~\textit{cartel membership}). The important entities in intelligence reports then map to central nodes in the knowledge graph. We introduce a novel method that extracts fine-grained entities in a few-shot setting (few labeled examples), given limited resources available to label the frequently changing entity types that intelligence analysts are interested in. It outperforms other state-of-the-art methods. Next, we identify challenges facing previous evaluations of zero-shot (no labeled examples) methods for extracting relations, affecting the step of populating edges. Finally, we explore the utility of the pipeline: given the goal of identifying important entities, we evaluate the impact of relation extraction errors on the identification of central nodes in several real and synthetic networks. The impact of these errors varies significantly by graph topology, suggesting that confidence in measurements based on automatically extracted relations should depend on observed network features.
\end{abstract}
\input{paper_body}


%
%

\bibliographystyle{splncs04}
\bibliography{ner_and_graphs}






\section{Appendix}
\subsection{Detailed NER results}
\label{s:detail}

Tables~\ref{table:nerbuilding}, \ref{table:nerlocation}, and \ref{table:nerorganization} contain detailed NER results for fine-grained entity types within coarse-grained entity types \textsc{Building} (\ref{table:nerbuilding}) and \textsc{Organization} (\ref{table:nerorganization}), and for \textsc{GPE} (\ref{table:nerlocation}).

\begin{table}
\begin{center}
{\small
\begin{tabular}{||c | c | c | c | c | c | c | c | c||} 
 \hline
  &  & hotel & library & restaurant & sportsfacility & hospital & theater & airport \\ [0.5ex] 
 \hline\hline
 \multirow{3}{*}{CONT} & precision & 42.9 & 46.6 & 49.0 & 52.4 & 58.8 & 48.1 & 57.1 \\
  & recall & 38.1 & 43.8 & 43.1 & 45.8 & 52.1 & 42.8 & 50.6  \\
  & F1 & 40.3 & 44.9 & 45.7 & 48.7 & 54.9 & 44.9 & 53.5 \\
 \hline
 \multirow{3}{*}{LTP} & precision & 50.6 & 56.1 & 47.1 & 67.4 & 63.4 & 58.5 & 75.8 \\
  & recall & 59.7 & 67.5 & 49.9 & 76.7 & 69.6 & 62.0 & 73.2  \\
  & F1 & 54.8 & 61.3 & 48.4 & 71.7 & 66.3 & 60.1 & 74.4 \\
 \hline 
 \multirow{3}{*}{Ours} & precision & 79.2 & 80.4 & 59.9 & 100 & 67.5 & 72.0 & 66.4 \\ 
  & recall & 74.7 & 70.3 & 45.7 & 62.1 & 77.7 & 84.2 & 86.8 \\
  & F1 & 76.9 & 75.0 & 51.8 & 76.6 & 72.2 & 77.6 & 75.2 \\[1ex] 
 \hline
\end{tabular}}

\end{center}
\caption{Results on fine-grained entity types within the \textsc{Building} coarse-grained type.}
\label{table:nerbuilding}
\end{table}

\begin{table}
\begin{center}
\begin{tabular}{||c | c | c||} 
 \hline
  &  & gpe \\ [0.5ex] 
 \hline\hline
 \multirow{3}{*}{CONT} & precision & 52.6 \\
  & recall & 35.7 \\
  & F1 & 42.4 \\
 \hline
 \multirow{3}{*}{LTP} & precision & 65.1 \\
  & recall & 65.0 \\
  & F1 & 65.0 \\
 \hline
 \multirow{3}{*}{LTP\textunderscore{N}} & precision & 59.1  \\ 
  & recall & 74.6  \\
  & F1 & 65.9\\
 \hline 
 \multirow{3}{*}{Ours} & precision & 73.4 \\ 
  & recall & 72.2 \\
  & F1 & 72.8 \\[1ex] 
 \hline
\end{tabular}

\end{center}
\caption{Results for the \textsc{GPE} fine-grained entity type.}
\label{table:nerlocation}
\end{table}

\begin{table}
\begin{center}
{\footnotesize
\begin{tabular}{||c | c | c | c | c | c | c | c | c | c | c ||} 
 \hline
  &  & showorg & comp & media & polparty & edu & spteam & govt & religion & spleague \\ [0.5ex] 
 \hline\hline
 \multirow{3}{*}{CONT} & precision & 44.9 & 43.4 & 46.6 & 49.5 & 53.6 & 54.2 & 37.9 & 44.3 & 49.9 \\
  & recall & 35.3 & 32.4 & 38.7 & 40.7 & 53.0 & 45.3 & 31.9 & 37.0 & 44.1 \\
  & F1 & 39.1 & 36.9 & 42.0 & 44.5 & 52.9 & 49.3 & 34.3 & 40.1 & 46.6 \\
 \hline
 \multirow{3}{*}{LTP} & precision & 52.0 & 59.6 & 56.4 & 61.5 & 54.7 & 64.2 & 38.0 & 47.9 & 55.5 \\
  & recall & 43.9 & 52.4 & 52.4 & 63.8 & 53.4 & 60.2 & 38.6 & 53.7 & 62.4 \\
  & F1 & 47.5 & 55.6 & 54.3 & 62.6 & 54.0 & 62.0 & 38.1 & 50.5 & 58.8 \\
 \hline
 \multirow{3}{*}{LTP\textunderscore{N}} & precision & 48.6 & 56.8 & 57.1 & 57.7 & 56.6 & 62.8 & 35.3 & 46.5 & 48.5 \\ 
  & recall & 52.6 & 63.3 & 66.2 & 67.3 & 68.2 & 70.3 & 45.3 & 57.6 & 68.2\\
  & F1 & 50.5 & 59.8 & 61.2 & 62.1 & 61.8 & 66.3 & 39.6 & 51.4 & 56.5 \\
 \hline 
 \multirow{3}{*}{Ours} & precision & 100 & 64.4 & 65.5 & 100 & 66.4 & 100 & 62.9 & 64.6 & 100 \\ 
  & recall & 1.0 & 68.7 & 72.3 & 78.1 & 71.0 & 72.1 & 41.2 & 71.2 & 54.9 \\
  & F1 & 2.0 & 64.5 & 68.7 & 87.7 & 68.6 & 83.8 & 49.8 & 67.7 & 70.9 \\[1ex] 
 \hline
\end{tabular}}

\end{center}
\caption{Results on fine-grained entity types within the \textsc{Organization} coarse-grained type.}
\label{table:nerorganization}
\end{table}
\end{document}

%% file: paper_body.tex
\section{Introduction}
\label{s:intro}

\begin{figure}[t]
    \centering
    \includegraphics[scale=.43]{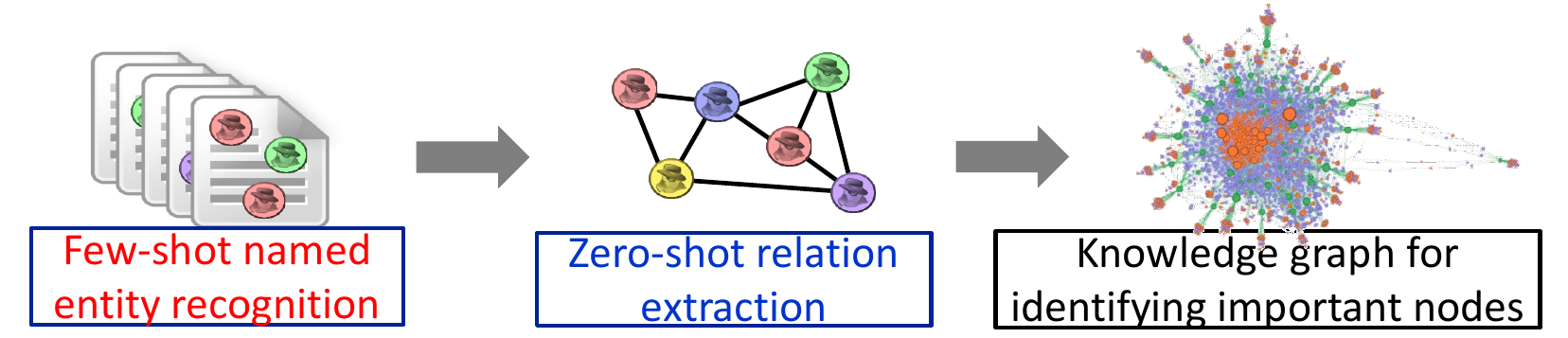}
    \caption{The pipeline from text to knowledge graph network, which can be used to identify important entities.}
    \label{fig:intro}
\end{figure}

Given intelligence reports, analysts often need to identify important entities based on their relationships with other entities. A simple pipeline for this involves populating a knowledge graph by extracting entities from text, which correspond to nodes, as well as relations, which correspond to edges. Identifying important entities then becomes a problem of identifying central nodes in the knowledge graph as in Figure~\ref{fig:intro}. However, constructing such a graph is challenging because entity and relation types of interest change frequently and entity types are \textit{fine-grained}, e.g., terrorists or cartels, within coarse-grained entity categories of people, organizations, and locations. 
The types of relations (e.g.~\textit{cartel membership}) extracted also change frequently. To adapt to the changing requirements, and given limited resources for labeling informative examples of new entity and relation types, our methods for extracting entities and relations need to operate in a few-shot or zero-shot setting (i.e., relying on few or no labeled examples).



We introduce a few-shot method for extracting \textit{fine-grained} named entities to serve as nodes in the knowledge graph. 
Using the limited labeled examples, our approach selects probability thresholds and optimal keywords out of a set of keyword candidates produced by a generative language model. Then, using the selected thresholds and keywords, our approach adapts text entailment, which outputs a probability that one text span (\textit{premise}) implies another (\textit{hypothesis}), to classify fine-grained entities. It outperforms state-of-the-art approaches such as Least Token Probability~\cite{liu2020ltp}, which is an active learning approach, and CONTaiNER~\cite{das2021container}, which is a few-shot learning approach using contrastive learning. 

For the step of populating knowledge graph edges, we identify challenges in the evaluation of many recent zero-shot relation extraction methods~\cite{chen-li-2021-zs,tran-etal-2022-improving,najafi-fyshe-2023-weakly}, where we find datasets~\cite{han-etal-2018-fewrel, gao-etal-2019-fewrel,chen-li-2021-zs} have significant missing annotations. We focus on zero-shot settings since examples of relations in text may be difficult to find. 

Finally, we assess how errors in the knowledge graph populated using these low resource information extraction methods affect identification of important entities.
Specifically, we compare how central nodes on the learned graph differ from those on a ground truth graph for various relations. Since typical evaluation datasets do not have ground truth knowledge graphs, we consider different types of potential ground truth graphs, both real and synthetic.

In summary, to identify important entities in intelligence report text, we: 

1. \textit{Define a knowledge graph construction pipeline}, specifying nodes as extracted fine-grained entities from text, and specifying edges through relation extraction between entities in a low-resource setting. 

2. \textit{Introduce a novel few-shot method} for extracting fine grained entities that outperforms other state-of-the-art approaches.

3. \textit{Identify challenges} for evaluation of zero-shot relation extraction methods.

4. \textit{Analyze the effect of errors in the knowledge graph}, finding that effects of relation extraction performance significantly depend on the graph topology.


\section{Task Definition}
\label{s:task}
We introduce the tasks in the pipeline:

\begin{figure}[t]
    \centering
    \includegraphics[scale=.278]{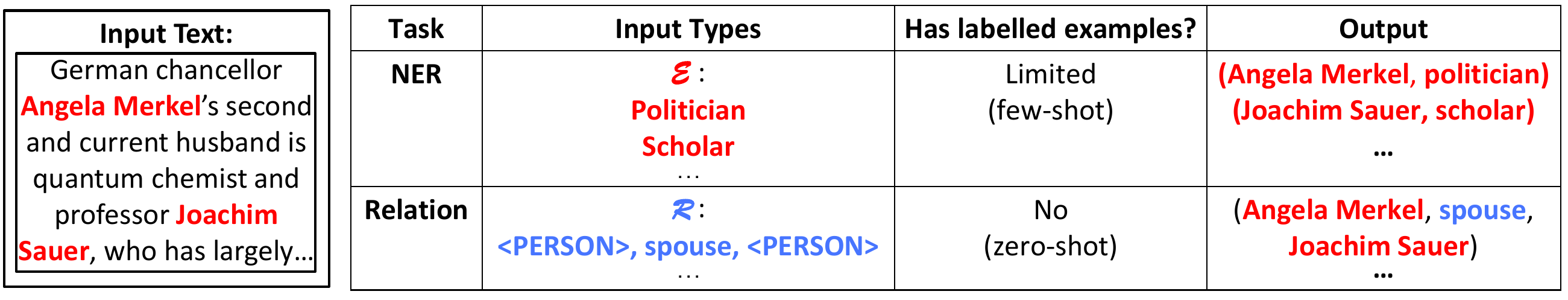}
    \caption{Example input and output of few-shot fine-grained NER (first row), the first step, and of zero-shot relation extraction (second row), the second step.}
    \label{fig:nerinputoutput}
\end{figure}

\textbf{Few-shot fine-grained named entity recognition.}  To define nodes of the knowledge graph, we perform a \textit{few-shot fine-grained named entity recognition} (NER) task to extract entities that belong to fine-grained types (e.g.~\textit{Politician}) within \textsc{person}, \textsc{organization}, and \textsc{location} coarse-grained entity types, which involve humans and therefore are of interest to intelligence analysts and social scientists. The inputs are a set of entity types $\mathcal{E}$, text from which to extract and classify entities, and few labeled examples of each entity type. Figure~\ref{fig:nerinputoutput} shows the output of sentence-level sets of entity-type tuples $\langle a,A\rangle$ where $a \in A \in \mathcal{E}$.


\textbf{Zero-shot relation extraction.} To define edges of the knowledge graph, we perform the \textit{zero-shot relation extraction} task to extract relations among entities belonging to coarse-grained types \textsc{person} and \textsc{organization}. The inputs are a set $\mathcal{R}$ of templates each containing a relation type and a pair of coarse- or fine-grained entity types, and text from which to extract and classify relations. Output are relations $\langle a,r,b\rangle$ fitting a template in $\mathcal{R}$, where $a,b$ are entities of a specified type (e.g.~\textsc{Person}) and $r$ is a relation type (e.g.~\textit{spouse}) as in Figure~\ref{fig:nerinputoutput}.

 \textbf{Identifying important entities.} Given the knowledge graph as input, the last step is to identify central nodes using a centrality or node ranking metric, where central nodes map to important entities. The output is an ordered list of the graph's top $k$ most important nodes. 



\begin{figure}[t]
    \centering
    \includegraphics[scale=.335]{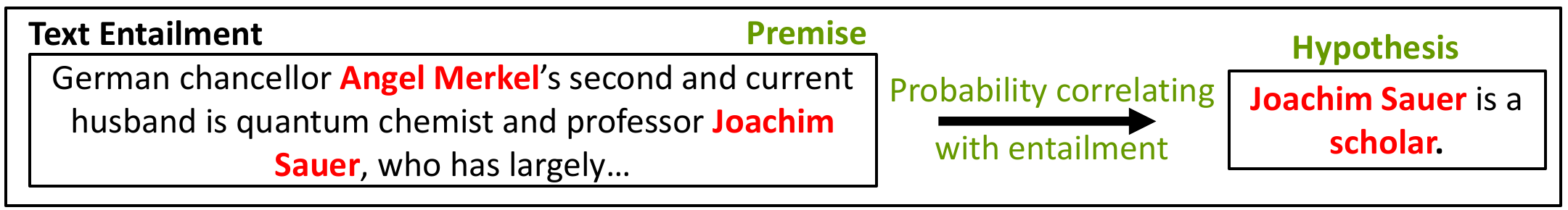}
    \caption{Example of text entailment for fine-grained NER.}
    \label{fig:te}
\end{figure}

\begin{figure}[t]
    \centering
    \includegraphics[scale=.285]{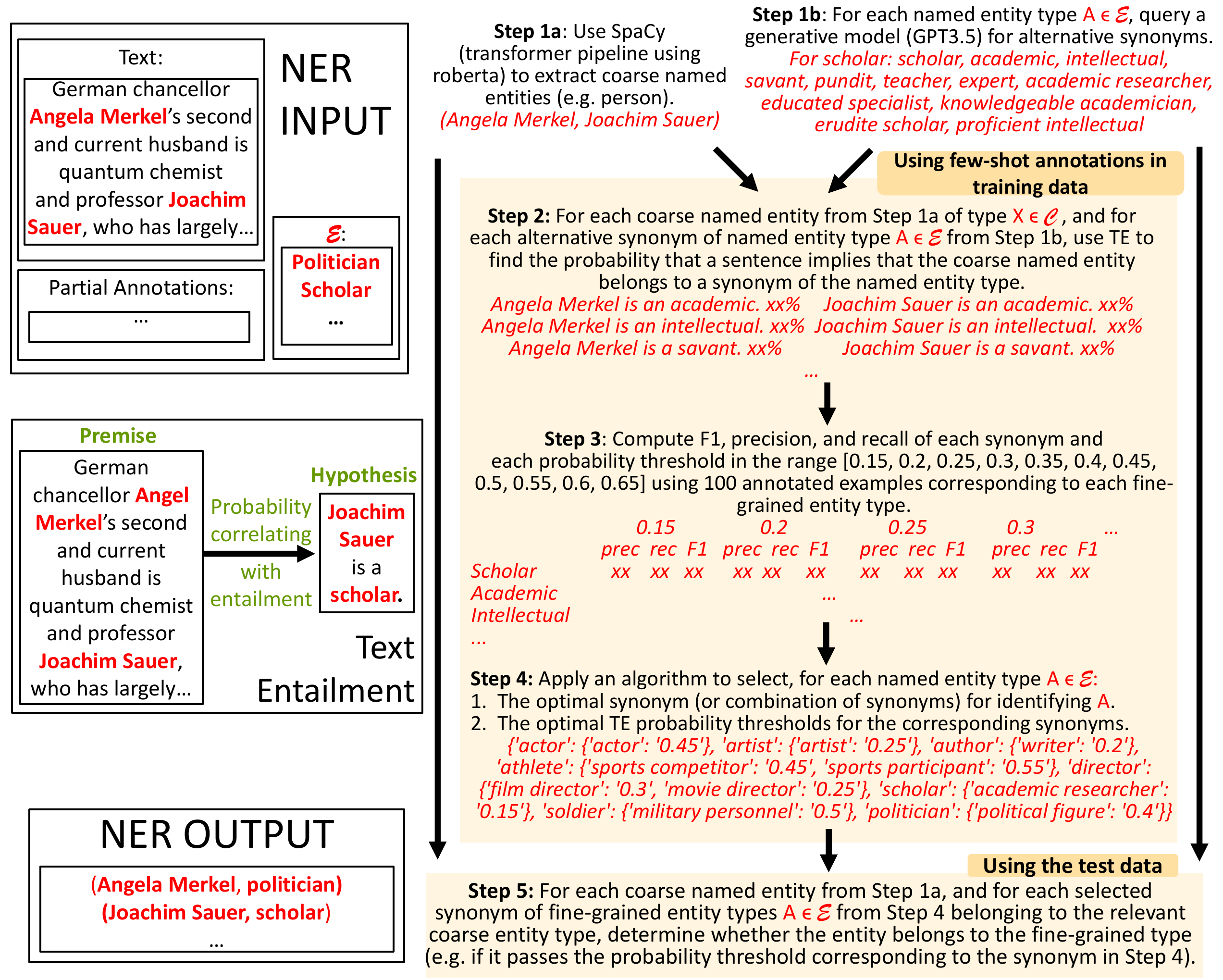}
    \caption{Steps of the NER approach, with examples in red.}
    \label{fig:nermethod}
\end{figure}

\section{Method for Few-shot Fine-grained Named Entity Recognition}
\label{s:nermethod}

We introduce a few-shot fine-grained named entity extraction method for populating the knowledge graph, which uses text entailment as in Figure~\ref{fig:nermethod} to determine if a coarse-grained entity in text belongs to a fine-grained entity type.

\textbf{Text entailment.} \textit{Text entailment} (TE), also known as natural language inference (NLI), takes two text spans, a \textit{premise} and \textit{hypothesis} as input and outputs a probability that the premise entails the hypothesis as in Figure~\ref{fig:te}.


\textbf{Notation.} We refer to the set of coarse-grained entity types (e.g.~\textsc{Person}, \textsc{Organization}) as $\mathcal{C}$, and of fine-grained entity types (e.g.~\textit{Politician}) as $\mathcal{E}$.

\textbf{Models.} The method uses (1) a SpaCy transformer model to extract coarse-grained entities from text, having strong zero-shot performance; (2) a generative model \texttt{text-davinci-003} as a lexical resource to produce single and multi-word synonyms of each fine-grained entity type; (3) \texttt{deberta-v2-xxlarge}, 1.5B parameters, fine-tuned on mNLI (\cite{williams2018broad}; labeled with TE examples) for TE (\S\ref{s:task}).

\textbf{Training on limited examples.} As in Figure~\ref{fig:nermethod}, using positive and negative labeled examples, our method: (A) Selects \textit{keyword representations} $K_A$ (e.g.~\textit{scholar, academic, teacher}) for each fine-grained entity type $A$ (e.g.~\textit{Scholar}) to use for the TE hypotheses; (B) Selects a \textit{probability threshold} for each keyword, to apply to the TE probability output to distinguish if a coarse-grained entity (e.g. \textit{Joachim Sauer}) belongs to a fine-grained entity type (e.g. \textit{Scholar}).

For each sentence, given sets $C_X$ of coarse-grained entities (e.g.~\textit{Joachim Sauer}) for each type $X \in \mathcal{C}$ (e.g.~\textsc{Person}) and sets $K_A$ of synonyms (e.g.~\textit{academic}) for each fine-grained entity type $A \in \mathcal{E}$ (e.g.~\textit{Scholar}), the method constructs \textit{hypotheses} for TE as sentences of the form $c_X$ \textit{is a/an} $k_A$ (e.g.~\textit{Joachim Sauer is a scholar}), where each $\langle c_X$, $k_A\rangle$ pair are from the Cartesian product $\{C_X \times K_A \mid A \textrm{ is of type } X\textrm{, } c_X \in C_X \textrm{ is a coarse entity, } k_A \in K_A \textrm{ is a synonym}\}$ as in step 2 of  Figure~\ref{fig:nermethod}. The \textit{premise} corresponding to these hypotheses is the sentence having the coarse-grained entities in sets $C_X$. From the probability outputs, for each synonym $k_A \in K_A$, the method (using the few labeled examples as ground truth) computes precision, recall and F1 scores to distinguish positive and negative instances of fine-grained entity types given probability thresholds in interval $[0.1,0.6]$ as in Step 3 of Figure~\ref{fig:nermethod}.

Finally, the method selects, for each synonym, probability thresholds that maximize F1 of classifying entity types over the limited labeled examples. For each fine-grained entity type, it selects at most three synonyms that maximize F1, filtering out synonyms that produce many false positives or low recall.

\textbf{Applying the method.} The method then applies the selected probability thresholds for each fine-grained named entity type and synonyms from the few-shot examples on the data. It identifies a positive instance of a fine-grained entity type if the TE probability output for any of the selected synonyms (1) exceeds its corresponding probability threshold, and (2) is maximum out of the probabilities corresponding to all synonyms that exceed their corresponding threshold. 

\section{Results for Few-shot Fine-grained Named Entity Recognition}
\label{s:nereval}

\textbf{Few-NERD dataset~\cite{ding-etal-2021-nerd}.} To evaluate our few-shot fine-grained named entity recognition approach, we use the manually annotated Few-NERD dataset  containing 66 fine-grained entity types, where the training set is $\sim$150K sentences, and the test set is $\sim$40K sentences, all from Wikipedia. Our evaluation is on all fine-grained entities within coarse-grained types \textsc{person}, \textsc{organization}, and \textsc{building} (i.e., location that humans occupy) in the dataset, and for the \textsc{gpe} type ($23$ types total). These are relevant for the intelligence analysts' application, as well as for various other social science applications, as entity types involving humans. Unfortunately, recent few-shot fine-grained NER methods struggle to perform well on Few-NERD~\cite{das2021container}. 

\textbf{Evaluation setting.} Our evaluation uses 100 manually annotated instances for each fine-grained named entity type, over \textit{all} fine-grained entity types within coarse-grained types of \textsc{Person}, \textsc{Organization}, and \textsc{Building}, as well as GPE. We chose 100 as a realistic number of fine-grained entities to acquire. 
Our evaluation setting differs from other few-shot evaluations which randomly sample a few fine-grained entity types and use fewer labeled examples; we adapt our setting for recent state-of-the-art approaches.

Since Few-NERD contains many well-represented fine-grained entities, our use case scenario of few-shot in-domain NER results in very imbalanced datasets. We sub-sample the data using the guideline in \cite{mayhew2019named} regarding the stability of ratio of entity tokens to total tokens in NER datasets, with a ratio of $\sim$0.04. We keep all sentences containing the fine-grained entity of interest and randomly sample the remaining sentences until reaching the desired token ratio. %


\textbf{Baselines.} 
There are two main relevant approaches for fine-grained NER: active learning (AL) and few-shot learning. For the AL baseline, we use the state-of-the-art Lowest Token Probability (LTP) approach~\cite{ liu2020ltp}, which is a strategy for Conditional Random Field-based models, combining global (sentence level) and local (token level) information to identify the most informative samples in a dataset. 
The original LTP algorithm samples a static number of sentences with each AL iteration, making no guarantees about the number of entities within the selected sentences. Instead, in our experiments, we selected samples ordered by their LTP score until our AL iteration entity count budget was reached. To reach our budget of 100 manually labeled entities, we sampled data over 4 AL iterations, targeting a selection of 25 manually annotated entities in each. 

For each fine-grained entity type, we tested LTP model performance using two versions of the Few-NERD dataset as training data.  The first version contained only the single fine-grained entity of interest; all other entity types were replaced with the non-entity token. The second version of the dataset, which we refer to as LTP with negative samples (LTP\textunderscore{N}), augments the single entity version of the dataset by adding 16 coarse-grained entity types extracted using the SpaCy transformer model. 
All of our AL results were averaged over 4 runs.

For a few-shot learning baseline, we used CONTaiNER~\cite{das2021container}, a state-of-the-art approach that uses contrastive learning, optimizing an objective differentiating between entity categories based on their Gaussian-distributed embeddings. As with LTP\_N, we use SpaCy to label negative samples in our experiments. 




\begin{table}[t]
\begin{center}
{\footnotesize
\begin{tabular}{||c | c | c | c | c | c | c | c | c||} 
 \hline
  &  & artist/ & \multirow{2}{*}{soldier} & \multirow{2}{*}{athlete} & \multirow{2}{*}{polit.} & \multirow{2}{*}{director} & \multirow{2}{*}{actor} & \multirow{2}{*}{scholar} \\ 
   &  & author & &  &  & &  & \\ [0.5ex] 
 \hline\hline
 \multirow{3}{*}{CONT} & precision & 61.7 & 56.4 & 59.7 & 56.5 & 60.0 & 70.6 & 56.6  \\  
  & recall & 55.6 & 43.0 & 43.9 & 43.6 & 47.0 & 62.7 & 42.0 \\
  & F1 & 58.3 & 48.8 & 50.5 & 49.1 & 52.6 & 66.2 & 48.1 \\
 \hline
 \multirow{3}{*}{LTP} & precision & 45.1 & 56.9 & 77.7 & 59.7 & 54.9 & 74.7 & 58.4 \\
  & recall & 28.8 & 54.8 & 72.0 & 45.6 & 60.0 & 65.9 & 55.3  \\
  & F1 & 34.9 & 55.6 & 74.7 & 51.7 & 57.3 & 69.9 & 56.2 \\
 \hline
 \multirow{3}{*}{LTP\textunderscore{N}} & precision & 47.0 & 56.5 & 70.1 & 55.0 & 56.5 & 68.7 & 53.7  \\ 
  & recall & 41.0 & 64.9 & 81.1 & 63.3 & 64.4 & 74.4 & 68.7 \\
  & F1 & 43.2 & 60.2 & 75.2 & 58.4 & 60.1 & 71.4 & 60.2 \\
 \hline
 \multirow{3}{*}{Ours} & precision & 79.7 & 84.2 & 88.4 & 78.9 & 89.8 & 79.9 & 82.0 \\ 
  & recall & 72.5 & 66.9 & 88.3 & 75.0 & 78.2 & 78.9 & 50.3 \\
  & F1 & 75.9 & 74.6 & 88.3 & 76.9 & 83.6 & 79.4 & 62.4 \\
 \hline
\end{tabular}}
\quad
{\footnotesize
\begin{tabular}{||c ||} 
 \hline
  Avg over \\ 
   23 types\\[0.5ex] 
 \hline\hline
  52.2 \\ 
  43.7 \\
  47.3 \\
 \hline
 58.4\\
 57.8\\
 57.7\\
 \hline
 55.2\\ 
 65.3\\
 59.7\\
 \hline
 79.4\\ 
 67.3\\
  70.2\\
 \hline
\end{tabular}}

\end{center}
\caption{\textbf{Left}: Percentage point results on fine-grained entity types within the \textsc{Person} coarse-grained type. \textbf{Right}: Average percentage point results over all fine-grained entity types within \textsc{Person}, \textsc{Organization}, \textsc{Building}, and \textsc{gpe}. In addition to the \textsc{Person} fine-grained types in Table~\ref{table:nerperson}, the types are \textit{show organization, company, media, political party, education, sports team, government, religion, sports league, 
hotel, library, restaurant, sports facility, hospital, theater, airport,} and \textit{GPE}. }
\label{table:nerperson}
\end{table}



%

\textbf{Results.} We provide the results of our approach compared to the baselines on fine-grained entity types within the coarse-grained entity type \textsc{Person} in Table~\ref{table:nerperson} (left) and provide average results over all 23 fine-grained entity types in \textsc{Person, Organization, Building} coarse-grained types and \textsc{GPE} in Table~\ref{table:nerperson} (right). Further result details are in Tables~\ref{table:nerbuilding}, \ref{table:nerlocation}, and \ref{table:nerorganization} of \S\ref{s:detail}. For baselines, for each fine-grained entity type, the few manually labeled examples are all positive. Our approach uses half positive and half negative examples that still share the same coarse-grained entity type. 
We also experimented with open-source Alpaca and Llama \cite{touvron2023llama}, converting TE hypotheses to questions such as \textit{Is [coarse-entity] a [fine-grained-entity-type]?}, but observed inconsistent formatting issues and poorer performance.


\section{Relation Extraction for Knowledge Graph Construction}
\label{s:rel}

Next, we populate relations in a knowledge graph and explore how performance of this step affects identification of important entities.

\subsection{Challenges facing zero-shot relation extraction evaluation} 
\label{s:relchallenges}

The pipeline uses zero-shot relation extraction, which outputs relations $\langle a,r,b\rangle $ where $a$, $b$ are entities and $r$ is a relation type as in Figure~\ref{fig:nerinputoutput}. For a knowledge graph, this output specifies an edge to exist between nodes that correspond to entities $a$ and $b$. 
However, we identify several challenges with using popular zero-shot relation extraction datasets for evaluation and encourage addressing these to improve performance evaluations.

\textbf{Challenges of using FewRel~\cite{han-etal-2018-fewrel, gao-etal-2019-fewrel}.} FewRel uses Wikipedia as the corpus and Wikidata as the knowledge base for annotation, with further filtering by crowdworker annotators. It consists of 80 relations, where each relation has 700 examples. It is designed for few-shot relation extraction evaluation, for computing accuracy for a set of randomly selected relations. To evaluate \textit{zero-shot} relation extraction methods, \cite{chen-li-2021-zs} discusses an evaluation strategy that instead computes precision and recall, where \cite{tran-etal-2022-improving,najafi-fyshe-2023-weakly} follow. However, FewRel only annotates a single relation for each sentence, increasing the likelihood that an evaluation approach incorrectly identifies extracted relations as false positives (e.g.~the \textit{mother} relation implicitly involving the \textit{child} relation). 


\textbf{Challenges of using WikiZSL~\cite{chen-li-2021-zs}.} WikiZSL is a subset of the Wiki-KB dataset with 94,383 instances, and is generated with distant supervision. Entities come from Wikipedia articles and are linked to the Wikidata knowledge base to get relations, and each sentence has one or more relation annotations. However, it misses many annotations, leading evaluations to incorrectly count false positives. One factor could be that the annotations are not filtered by human annotators. We observe, for example, that 1070 unique sentences have a \textsc{father} relation annotation, but only three of these include a \textsc{child} relation. 

\subsection{Naive zero-shot relation extraction evaluation}
\label{s:releval}

\begin{figure}[t]
    \centering
    \includegraphics[scale=.33]{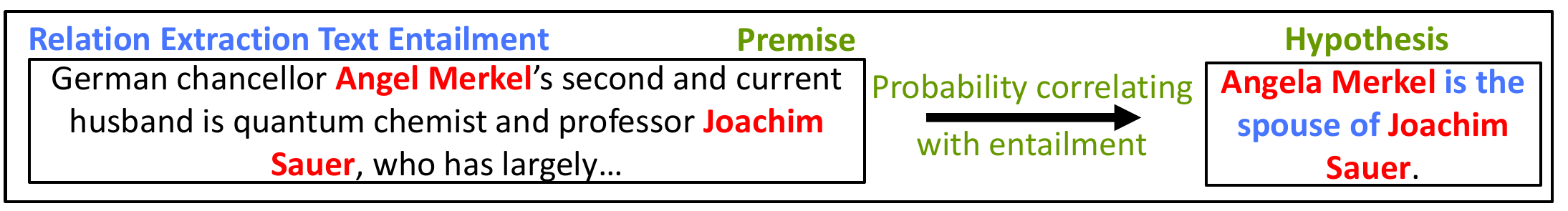}
    \caption{Example of text entailment for zero-shot relation extraction.}
    \label{fig:relmethod}
\end{figure}

To observe how relation extraction performance could affect identification of central nodes in \S\ref{s:errorprop}, we demonstrate a naive zero-shot relation extraction approach for populating a knowledge graph using TE (\texttt{deberta-v2-xxlarge} fine-tuned on mNLI) on relations among entity types of \textsc{Person} and \textsc{Organization}. These types, involving humans, are the ones that typically interest intelligence analysts and social scientists. 
If a sentence containing a pair of entities (e.g. \textit{Joachim Sauer, Angela Merkel}) implies a hypothesis concatenating the entities and relation type (e.g. \textit{Angela Merkel is the spouse of Joachim Sauer}; Figure~\ref{fig:relmethod}) with probability $0.8$ or greater, then the naive method concludes a positive instance of a relation type. Our evaluation follows \cite{chen-li-2021-zs,tran-etal-2022-improving,najafi-fyshe-2023-weakly}, where we select $10$ random relations as unseen and random train/test splits for each relation, dropping the training data. In Table~\ref{table:rel}, we also share published performance of recent approaches for reference, but note that other evaluations do not restrict entity types of relations to belong to \textsc{Person} and \textsc{Organization} types.

\begin{table}[t]
\begin{center}
{\footnotesize
\begin{tabular}{||c | c | c | c | c || c || c ||} 
 \hline
  & ESIM & CIM & ZS-Bert  & Discrim. & Naive \\ \hline
 pr & 42.9  & 47.4 & 56.9 & 64.4 & 60.9\\ 
 \hline
 re &  44.2 & 49.1  &57.6 &62.6 & 64.4\\
 \hline
 F1 & 43.5 &48.2  & 57.3 & 63.5& 62.2\\

 \hline
\end{tabular}}
 \quad
{\footnotesize\begin{tabular}{||c | c | c | c | c || c || c ||} 
 \hline
  & ESIM & CIM & ZS-Bert  & Discrim. & Naive\\ \hline
 pr & 44.1  & 46.5 &60.5 &71.6 & 53.7\\ 
 \hline
 re & 45.5 & 47.9  &61.0 & 64.7& 57.9\\
 \hline
 F1 & 44.8 & 45.6 & 60.7 &67.9 &  53.9\\

 \hline
\end{tabular}}
\end{center}
\caption{Naive zero-shot relation extraction evaluation on FewRel (left) and WikiZSL (right), averaged over $10$ random relation types to aid analysis in \S\ref{s:errorprop}. For reference, we include published performance of ZS-Bert~\cite{chen-li-2021-zs},  discriminative learning~\cite{tran-etal-2022-improving}, Conditioned Inference Model (CIM)~\cite{rocktaschel2016reasoning}, and Enhanced Sequential Inference Model (ESIM)~\cite{chen2017enhanced}.}
\label{table:rel}
\end{table}

\subsection{Impact of Relation Extraction on Centrality Scoring}
\label{s:errorprop}

While there are significant issues with zero-shot relation extraction (in particular the overreporting of false positives), we consider the published performance of these methods, as well as that of our naive method based on text entailment, as a conservative estimate of relation extraction in a true network, and analyze the associated impact of the errors on centrality estimation.

Since the FewRel and WikiZSL datasets do not have ground truth knowledge graphs, we simulate ground truth graphs and learned knowledge graphs after relation extraction. Next, we compare the central nodes in the two graphs, identifying them using centrality metrics, to assess the effect of relation extraction performance on identifying important entities.


\textbf{Simulating a ground truth graph, $G_r$}: We consider four synthetic graph topologies and four real networks as ground truth graphs. Synthetic graphs are Erd\H{o}s--R\'{e}nyi random graphs (ER), Barab\'{a}si--Albert preferential attachment graphs (BA), Watts--Strogatz small-world graphs (WS), and Lancichinetti--For\-tu\-na\-to--Radicchi (LFR) community detection benchmark graphs. In all cases, the synthetic graphs had 500 nodes and average degree of approximately 12. The real networks used are social connections in the Provisional Irish Republican Army (PIRA, 391 nodes, 864 edges)~\cite{Gill2014}, co-involvement in kidnapping incidents within the Abu Sayyaf group (ASG, 207 nodes, 2550 edges)~\cite{Gerdes2014}, collaborations among network scientists (NS, 379 nodes, 914 edges)~\cite{Newman2006}, and interactions between participants at the ACM Hypertext 2009 conference (HT, 113 nodes, 2196 edges)~\cite{Isella2011}. In all real graphs, the largest connected component is used, and we consider unweighted, undirected graphs without the possibility of multiple edges between the same nodes. The real or generated graph is considered the ground truth graph $G_r=(V, E_r)$ defined by a single relation of interest $r$.

\textbf{Simulating a graph learned after relation extraction, $G_r^\prime$}: We construct graph $G_r^\prime$ that simulates a graph learned from relation extraction. To do this, we modify $G_r$, keeping the proportion of edges that matches $r$'s recall. Edges to be kept are selected uniformly at random. For all node pairs that do not share edges, we add false edges with probability $\frac{1-\textrm{precision}}{\textrm{precision}}\cdot|E_r|\cdot\textrm{recall}/\binom{|V|}{2}$. The same probability is applied to all such node pairs. Thus the edges present in $G_r^\prime$ are consistent with the precision and recall rates corresponding to the relation from relation extraction. 

\textbf{Identifying central nodes in $G_r$ and $G_r^\prime$.} The next step after constructing $G_r^\prime$ from $G_r$ is to identify central nodes in each and to evaluate performance based on the number of common central nodes in both graphs. We consider four centrality metrics---betweenness, closeness, eigenvector, and degree centrality---for the analysis. We find the $k$ nodes with the highest centrality scores in both graphs and report the size of the overlap of the two sets.

\begin{figure}
    \centering
    Betweenness Centrality\\
    \includegraphics[width=\textwidth]{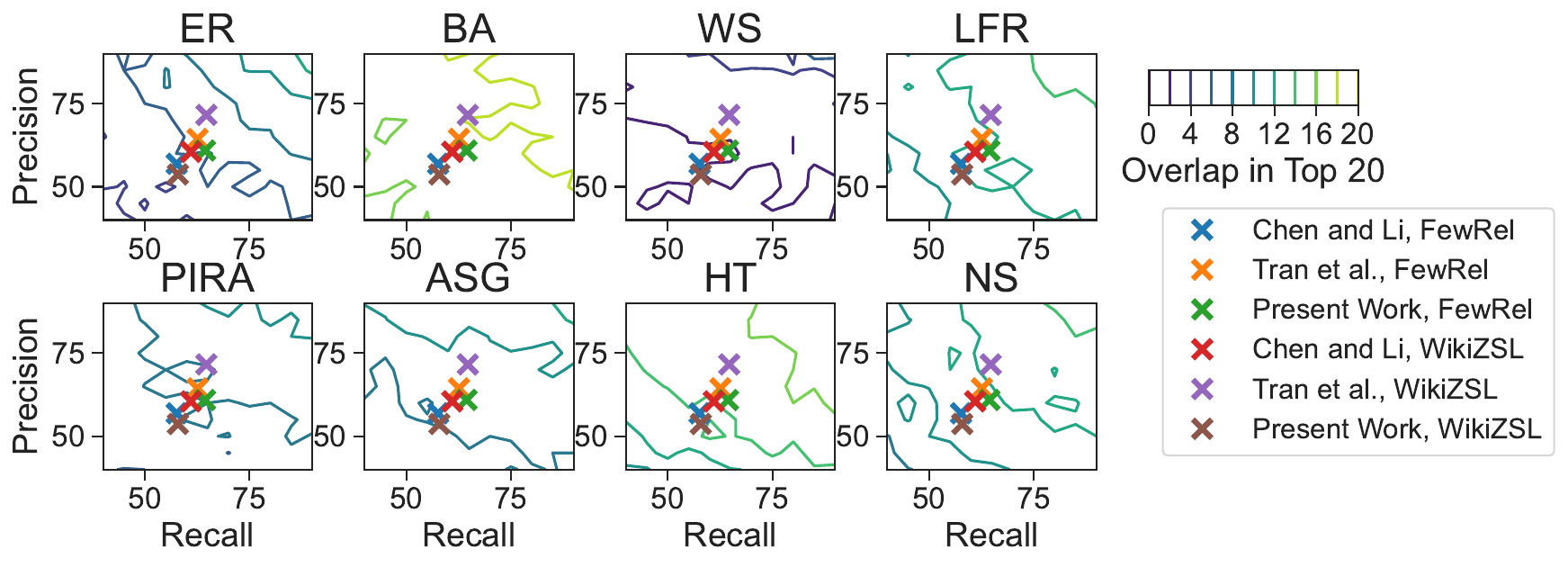}\\
    Closeness Centrality\\
    \includegraphics[width=\textwidth]{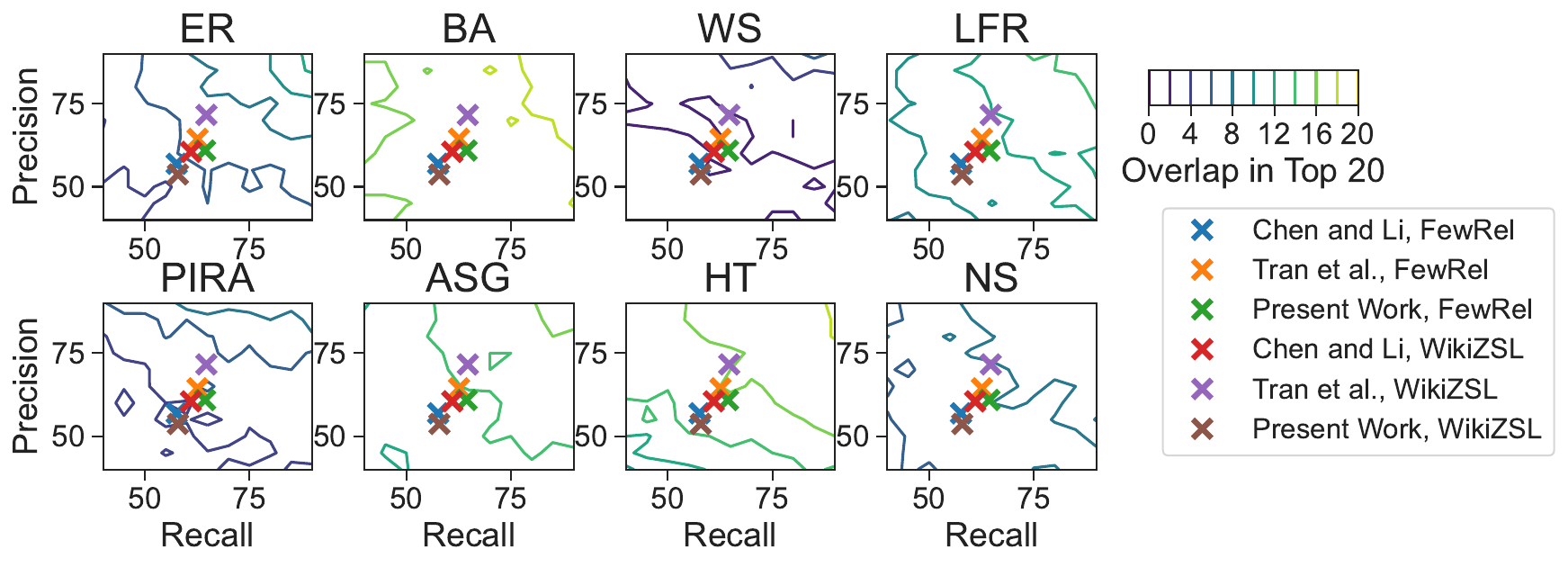}
    \caption{Plots of numbers of common nodes as a function of precision and recall for relation extraction. The size of the overlap between the 20 most central nodes in $G_r$ and $G_r^\prime$ is shown as a contour plot for all real and synthetic topologies. The $\times$ marks indicate the precision and recall of state-of-the-art relation extraction methods. The importance of relation extraction performance varies substantially from case to case: for example, performance is much better among the synthetic graphs with skewed degree distributions (BA and LFR), but central node estimation appears hindered by community structure (LFR worse than BA, WS worse than ER).}
    \label{fig:vis1}
\end{figure}

\textbf{Results.} We simulated relation extraction errors on the eight networks discussed in \S\ref{s:errorprop} and compared the $k$ most central nodes for $k\in\{5, 10, 20\}$. The largest errors occurred with betweenness and closeness centrality, and these results are illustrated in Figure~\ref{fig:vis1}. We present results for $k=20$; the others are similar. In addition to showing the overlap among the $k$ most central nodes in $G_r$ and $G_r^\prime$, we show where the methods reported in Table~\ref{table:rel} fall in this space (using precision and recall averaged over 10 random relation types). 
As we may expect, there is not much overlap between central nodes with ER and WS graphs, which have fairly homogeneous topologies and do not have stable important nodes. LFR graphs have greater overlap, and BA graphs, which are highly dependent upon their hubs, remain very consistent in the face of these errors. Performance with the real networks does not particularly resemble any of the synthetic data, but we see 
levels of overlap closer to the BA and LFR graphs, i.e., those with skewed degree distributions. Note the difference between HT and NS: HT is not very modular, while NS has clear community structure, and, similarly to the BA and LFR graphs, the less modular graph has greater overlap between the true and estimated central node sets. It seems that more modular structures include central nodes that are more likely to be hidden due to relation extraction errors, which is consistent with intuition: highly central nodes in graphs with skewed degree distributions and little clustering tend to be the high-degree nodes, which will likely have high degree even with errors in the data.


\section{Related Work}
\label{s:lit}
Knowledge graphs constructed from text are an effective tool for threat and risk analysis, and there are a number of works that present frameworks and algorithms for graph construction and analysis, for example \cite{wang2022evaluating, ren2022cskg4apt, jo2022vulcan,liu2022using, simek2018xlab}. In general, they use ontologies, often manually constructed, or deep learning models fine-tuned for specific domain, for NER and RE. 

For low-resource named entity recognition, there are algorithms for coarse-grained NER that perform well even in zero-shot settings provided that the data is well-structured such as the SpaCy model that we used, other BERT-based NER models~\cite{devlin2018bert} and  Stanford NER~\cite{manning2014stanford}. Supervised methods for fine-grained NER perform much better than low-resource methods~\cite{lothritz2020evaluating, xue2020coarse, leitner2019fine}. Besides targeting person, organization, and location fine-grained entities for the intelligence analysis applications, our approach differs from others by allowing $\sim$100 annotations, where other approaches mostly rely on either a much smaller selection of fine-grained entities with much fewer labeled examples for each~\cite{li2020few, huang2020few}, or on a more supervised setting with many more annotations available~\cite{radmard2021subsequence, zhou2021mtaal, siddhant2018deep, liu2020ltp}.

There is work using text entailment in various areas of information extraction such as event detection~\cite{lyu-etal-2021-zero} and relation extraction~\cite{rocktaschel2016reasoning,chen2017enhanced}. However, we propose a different way of using it where few-shot examples help select synonym(s) and probability thresholds, and this works well for the fine-grained NER task.


\section{Conclusion}

We address a key data analysis task of identifying important entities from text corpora with limited annotation resources in a pipeline, using a knowledge graph as an intermediate data structure. 
Each step is a challenging and active area of research, and we present a novel few-shot approach for the first step of identifying fine-grained named entities from text that outperforms other state-of-the art methods. We identify challenges facing previous zero-shot relation extraction evaluation approaches, and analyze how errors in relation extraction affect identifying important entities for a variety of graph topologies. In future work, we plan to explore approaches to improve zero-shot relation extraction methods and evaluation resources and develop novel ways to identify important entities.